# Nonlinear optical data transformer for Machine Learning

Mustafa Yildirim[1], Ilker Oguz[1], Fabian Kaufmann[3], Marc Reig Escalé[3,4], Rachel Grange[3], Demetri Psaltis[2] and Christophe Moser[1]

[1] Laboratory of Applied Photonics Devices, Ecole Polytechnique Fédérale de Lausanne (EPFL), Switzerland

[2] Optics Laboratory, Ecole Polytechnique Fédérale de Lausanne (EPFL), Switzerland

[3] Optical Nanomaterial Group, Institute for Quantum Electronics, Department of Physics, ETH Zurich, Switzerland

[4] Versics AG, Auguste-Piccard-Hof 1, HPT building, 8093 Zurich, Switzerland

## Abstract

Modern machine learning models use an ever-increasing number of parameters to train (175 billion parameters for GPT-3) with large datasets to obtain better performance. Bigger is better has been the norm. Optical computing has been reawakened as a potential solution to large-scale computing through optical accelerators that carry out linear operations while reducing electrical power. However, to achieve efficient computing with light, creating and controlling nonlinearity optically rather than electronically remains a challenge. This study explores a reservoir computing (RC) approach whereby a 14 mm long few-mode waveguide in $LiNbO_3$ on insulator is used as a complex nonlinear optical processor. A dataset is encoded digitally on the spectrum of a femtosecond pulse which is then launched in the waveguide. The output spectrum depends nonlinearly on the input. We experimentally show that a simple digital linear classifier with 784 parameters using the output spectrum from the waveguide as input increased the classification accuracy of several databases compared to non-transformed data, approximately a 10%. In comparison, a deep digital neural network (NN) with 40'000 parameters was necessary to achieve the same accuracy. Reducing the number of parameters by a factor of ~50 illustrates that a compact optical RC approach can perform on par with a deep digital NN.

## Introduction

Matrices establish connections between neurons of NN, and for each neuron, the input vector is multiplied by the corresponding matrix, the so-called multiply and accumulate operation (MAC). Subsequently, the resulting vector is passed through a nonlinear activation function. The heavy computational load of these operations prompted the optical implementation of MAC operation, using integrated photonics or 2D light modulators in free space. Recent work[1] implemented a reconfigurable free space vector-matrix multiplier based on the Stanford architecture[2], using a digital micromirror device and spatial light modulator for input vector and matrix encoding, respectively. The main advantage of this technique is that the MAC calculation does not consume energy while light propagates through transparent, non-scattering optical components. Besides, since matrices are implemented with 2D light modulators, operations with a matrix size of 1000-by-1000 can be executed in a single pass. Thermo-optically controlled Mach-Zehnder interferometer (MZI) based on planar matrix-vector multiplication (MVM) architectures are also being realized[3] using silicon photonics technology. Since the MAC calculation takes place passively (via interference), the energy cost of MAC calculation is reduced compared to digital computers. This technology is on its way to being commercialized. However, thermal modulators consume energy constantly, which poses a scale-up problem. An optical crossbar array with a fix and forget phase change material (PCM) is introduced for MVM to reduce energy consumption, in which coupling ratios are set by PCM at crossing points[4]. Therefore, there is no need to power up the optical circuit continuously. Both techniques promise better energy efficiency for MAC calculation. Nonetheless, the growing number of parameters is a persistent challenge for scaling planar architectures.

In electronics, sending a vector through a neuron consists of five steps: reading the vector and matrix from memory, MAC calculation, sending the latter to a memory buffer, nonlinear activation calculation and writing the result to memory[5]. Among these steps, the most significant energy cost stems from memory accesses[6–8]. Data movement is generally the bottleneck for scaling rather than MAC calculation. Optical MVMs also suffer from memory accesses since the computation needs to be stored in memory before the non-linear activation function computation. However, related memory calls could be minimized with on-chip optical nonlinearity . Moreover, the addition of optical nonlinearity would allow the calculation of multiple DNN layers at once without changeover from optics to electronics for nonlinear activations. That would decrease memory access substantially.

Reservoir computing (RC) can reduce the number of parameters to learn[9]. Reservoirs are fixed complex passive, generally recurrent neural networks followed by a simple readout stage with few parameters to learn. The readout can be a simple linear regressor or classifier to be trained for a specific task. Therefore, the training cost of RC is low regardless of reservoir/task complexity. Expanding RC complexity does not increase the number of parameters or operations as in neural networks, making them appealing for scaling. With this motivation, an extensive repertoire of physical optical reservoir implementations has been realized based on semiconductor lasers, random diffusers, integrated waveguides, fibers[10–16]. The optical Kerr effect was recently used in multimode fibers to implement a reservoir for performing several machine learning tasks[17]. The work in ref[16] uses single-mode fibers with data encoded pulses for nonlinearity generation using $\chi^{(3)}$ effect. In the latter, nonlinearity occurs during propagation of the light pulse inside the fiber without optoelectronic conversion to achieve nonlinearity as opposed to refs[11,14]. That allows higher energy efficiency due to the absence of optoelectronic conversion. Although optical reservoirs present rich dynamics, either their large form factor or optoelectronic conversions make them expensive.

Integrated optical nonlinearities are needed to enable the potential of integrated linear accelerators and realize a compact on-chip RC approach. In this work, we introduce an on-chip optical nonlinearity for computation purposes and experimentally demonstrate its use for various classification tasks using the RC approach. We use a dispersion engineered Lithium niobate (LN) on insulator waveguide for all-optical nonlinearity realization on a chip. LN exhibits second and third-order nonlinearities because of its non-centrosymmetric structure. High refractive index contrast and the interplay between $\chi^{(2)}$ and $\chi^{(3)}$ of LN allow complex optical nonlinear processes at modest pump energies (tens of pJ pulse energies)[18,19]. We use a 14mm-long ridge waveguide that we pump with a data encoded femtosecond pulse. Data is encoded as the amplitude of the spectral components of the optical beam by a pulse shaper (see Methods). An overall experimental setup is shown in Figure-1a. These pulses experience nonlinear evolution ($\chi^{(2)}$ and

$\chi^{(3)}$ effects) through the waveguide, and a spectrometer records the corresponding spectrum at the output of the waveguide for each data sample. The output spectrum is then fed to a simple digital (i.e., non optical) ridge classifier. Note that the name ''ridge'' describing the classifier has no connection to the name of the waveguide (ridge).Linear discriminant analysis (LDA) is a technique to project high-dimensional datasets to low dimensions while preserving class diversity. We monitor LDA as an indicator for class separation. In our liver disease experiment (see Results for more detail), we observe that the nonlinear transformation of the input data by the LN waveguide expands the distance between the different classes and thus facilitates their classification. Figure-1b illustrates LDA of the liver disease dataset before and after the nonlinear transformation in the LN chip. Remarkably, the separation between healthy and other disease classes increases and better classification accuracy obtained. Optical transformation occurs in a planar waveguide, in view of a potential high level integration benefiting from on-chip lasers, modulators and detectors, our system is hereafter called intSOLO (integrated scalable optical learning operator[17]). In the remaining part of the paper, we analyze the nonlinear transformation performed by the intSOLO on various datasets in depth.

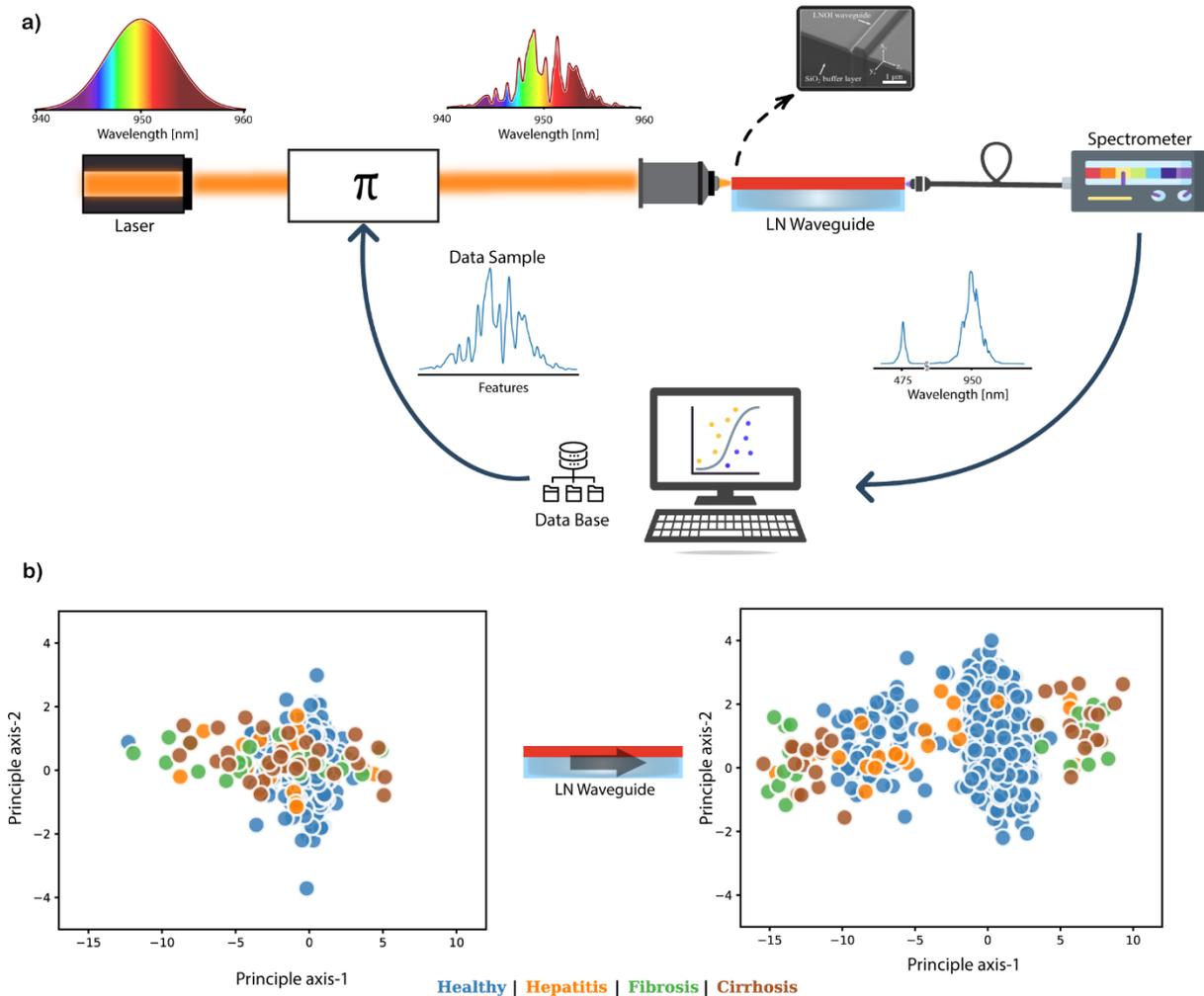

Figure 1. (a) Overall experimental setup. The spectral pulse shaper (π) encodes data samples to femtosecond pulse spectrum. The data pulses are focused in a lithium niobate on insulator ridge

waveguide. The pulse spectrum evolves nonlinearly through the 14 mm long waveguide. A spectrometer captures the output spectrum of each data instance. Further information processing steps are carried out on a digital computer (b) Liver disease dataset projection to two dimensions before and after the experiment, using the linear discriminant analysis method. The dataset has four classes: healthy, hepatitis, fibrosis, and cirrhosis. Each instance of the dataset is plotted by a painted dot, on which the color indicates which class the sample belongs.

## Results

### Vowel Classification

Spoken vowel classification is a generic task for machine learning systems. In this experiment, we use a dataset from ref[20], consisting of 12 formant frequencies extracted from audio recordings. The goal is to estimate a vowel from these frequencies. There are seven vowels and 273 samples in the dataset. The dataset is encoded as spectral amplitude by the pulse shaper (see Methods for details), and a corresponding spectrum for each instance is recorded after the waveguide. We discretize the output spectrum to produce a vector with 196 values. The spectrum is selected around the pump wavelength (950 nm) and the second harmonic (475 nm). To reveal the impact of the nonlinearity, we perform ridge classification on a test set of both original and transformed data through the LN waveguide. The ridge classifier is a single-layer linear mathematical method implemented digitally. Figure-2 shows confusion matrices calculated from the ridge model using original and transformed data (after the LN waveguide). The original data provides a test accuracy of 86.92%, whereas we obtain 96.06% classification accuracy with intSOLO. We attribute the improvement in performance to the nonlinear pulse propagation in the LN waveguide which increases the dimensionality of the output, serving to extract previously hidden data features. Since the ridge is a linear model, it guarantees that the accuracy improvement originates from the nonlinear propagation.

Additionally, we analyze the performance of intSOLO on several recognized classification techniques: support vector machines (SVM) and single hidden layer (200 neurons) neural network. A summary is shown in Table-1, where 'Baseline' means the original data used for the classification and 'LN' means the experimentally non-linearly optical transformed data. The LN transform increases the performance of linear methods (ridge and SVM linear). SVM polynomial for the LN case gives the worst performance, which means applying another nonlinear kernel to the transformed dataset scrambles it. The single hidden layer NN performs better with the original data than the transformed ones.

### Date Fruit Classification

We challenged intSOLO with another multiclass task having more input features and samples. There are 898 samples and seven different types of date fruits; the dataset is from ref[21]. In this dataset, each instance has 34 features (morphological features, shape etc.), which were extracted from date fruit images by computer vision techniques. All samples are transferred onto pulse spectra and propagated through the LN waveguide. The corresponding nonlinearly evolved spectrums are saved as in the vowel classification task. We performed a final linear ridge regression on raw and transformed datasets. Respective confusion matrices are depicted in Figure-2. The LN waveguide achieves a test accuracy of 96.75% compared with 90.04% test accuracy for the raw dataset provided. Different classifier methods and attained accuracy are presented in Table-1. All methods show a similar trend as in the previous case. Experimental SVM provides higher accuracy for the date fruit dataset.

### Liver Disease Dataset

A liver infection called hepatitis C can cause life-threatening health problems like cirrhosis or liver cancer in the later stage. It is crucial to diagnose early and get treatment. We use a dataset composed of blood samples categorized into healthy, hepatitis, fibrosis and cirrhosis people. Note that there are seven suspected samples in the dataset, they are considered as healthy for simplicity. The dataset is taken from ref[22] and includes 615 examples. Each instance has 10 parameters of blood, age and gender of participants. We obtained the nonlinearly transformed dataset as in the previous experiments. As shown in Figure-2, classification performance is raised by 11% compared to a raw dataset using the ridge classifier. To analyze further, we also compared accuracies with other decision layers. The results are listed in Table-1 show that higher accuracies are obtained with intSOLO. SVM polynomial provides better performance than SVM linear for the original dataset, but SVM linear with LN attains the best accuracy. It is thus natural to associate the non-linear transform of the waveguide as a sort of non-linear kernel for SVM.

### Numerical Studies

The simultaneous effect of second and third-order nonlinear effects could be interesting for all-optical information processing. Here, we observe both experimental and numerical input-output pairs, and we comment on possible reasons that enabled such data transformation. We numerically calculate pulse propagation using the generalized nonlinear Schrödinger equation via split step Fourier method, without second-order nonlinearity and Raman[23] effects. Experimentally measured pulse spectrums (before the LN waveguide) are used as numerical inputs. Figure-3 demonstrates corresponding simulated and measured output spectrums for modulated (Figure-3b) and unmodulated (Figure-3a) input, respectively. Self-phase modulation broadens the pulse spectrum, which is clearly seen for the experimental and numerical cases. There is a dip around 950 nm in the measured spectrum after the LN waveguide, which is known to disappear at low pump power[18]. It could be due to $\chi^{(2)}$ effect, whereby the second harmonic beam draws energy from the pump. Moreover, the back action of the second harmonic beam also produces frequencies near the pump wavelength[24]. Figure-3b shows the results of a data encoded pulse. Interestingly, the output has a peak around 950 nm, although there was not much power at that wavelength at the beginning. Overall, these results prove that the input-output relationship is nonlinear due to the interaction of material and fs pulses.

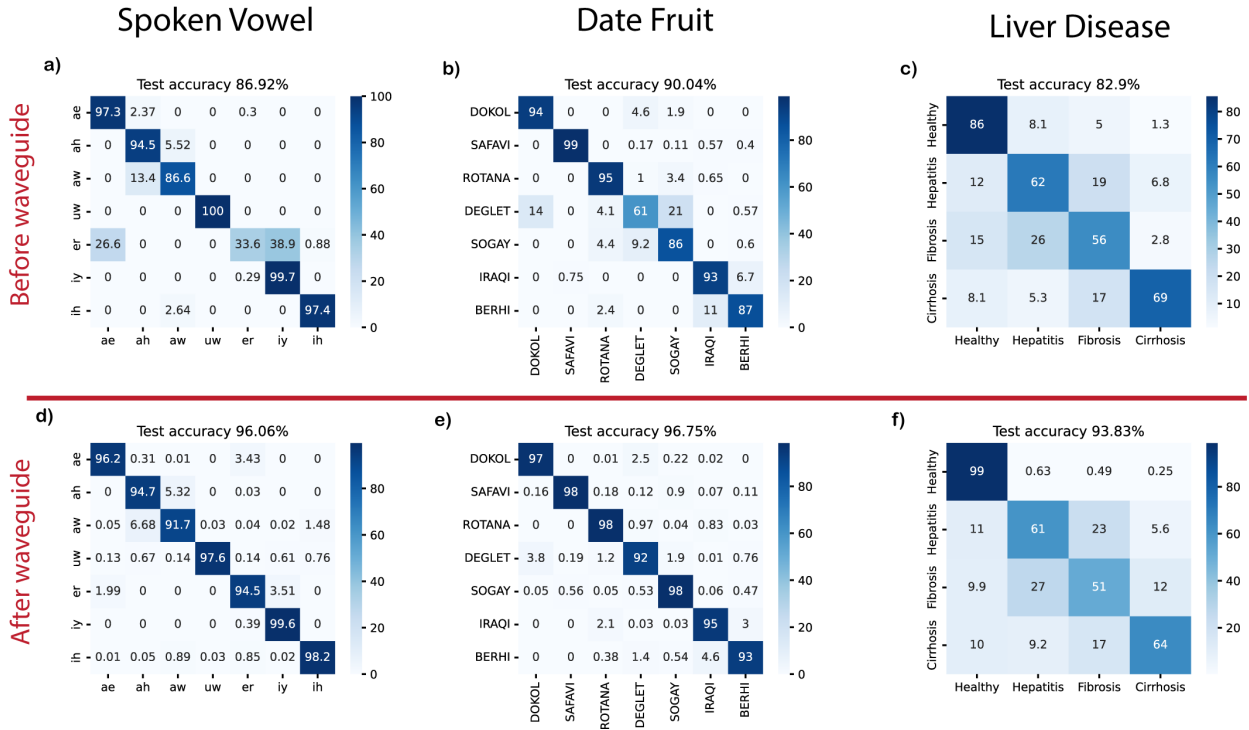

Figure 2. Confusion matrices of three experiments: spoken vowel (a, d), date fruit (b, e) and liver disease (c, f). The upper row (before waveguide) shows classification accuracies using the original dataset with the Ridge classifier (a, b, c). The bottom row (after waveguide) is calculated using nonlinearly transformed dataset (d, e, f). A Ridge classifier is also used for the original dataset.

| Methods (Parameter Counts) | Spoken Vowel | | Date Fruit | | Liver Disease | |
|---|---|---|---|---|---|---|
| | Baseline Acc(%) | LN Acc(%) | Baseline Acc(%) | LN Acc(%) | Baseline Acc(%) | LN Acc(%) |
| Ridge (784) | 86,92 | 96,06(2,22) | 90,04 | 96,75(1,18) | 82,90 | 93,83(1,59) |
| Neural Network (40000) | 97,61 | 94,49(2,73) | 90,80 | 88,64(2,40) | 91,81 | 93,02(2,00) |
| SVM Linear (784) | 94,60 | 95,62(2,24) | 90,64 | 91,95(2,17) | 87,81 | 93,23(1,87) |
| SVM Polynomial (784) | 94,65 | 65,80(5,27) | 85,72 | 79,04(2,74) | 90,34 | 82,64(6,6) |

Table 1: Classification accuracies (Acc) of three datasets for various digital classifier methods. Baseline acc means the original dataset is used to train a model. In the LN acc case, models are trained by an experimentally transformed dataset. Experiments are repeated 30 times, and the standard deviation of accuracies is written in parentheses. For all classification methods, the number of the model parameter is kept the same. SVM is used with polynomial and linear kernels. Numbers in parentheses next to methods indicate parameter counts of the method for the liver disease dataset.

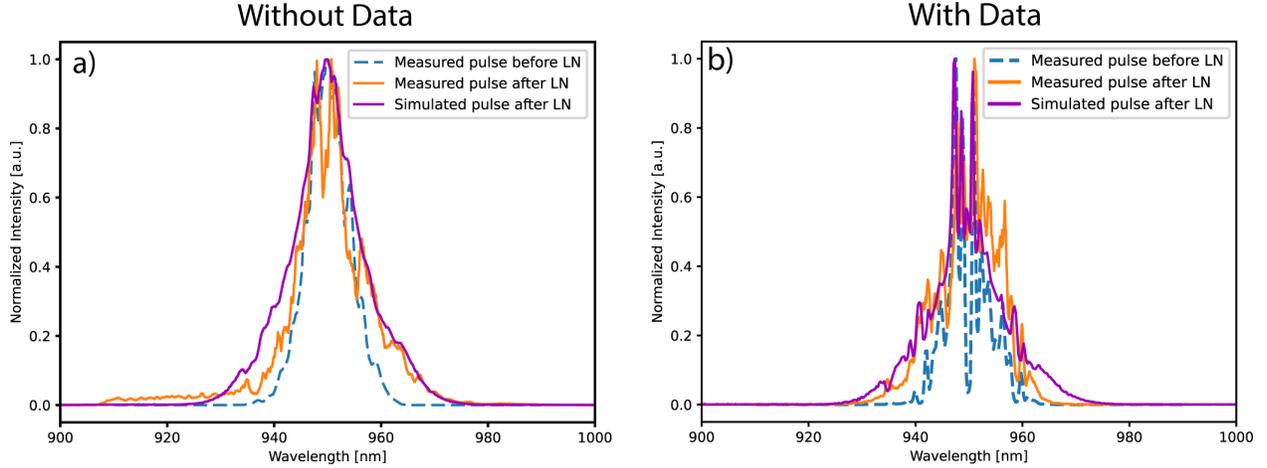

Figure 3: Input-Output pulse spectrums before and after LN waveguide with (b) and without (a) data encoding. Blue and orange plots are measured experimentally. A data sample having 13 features is encoded (b) Purple plots are obtained by propagation simulation of measured spectrums before LN.

## Discussion

We have shown that the nonlinearity in intSOLO provides a useful complex pre-processing before a simple digital linear classifier. Leveraging the LN waveguide as a frontend processing step substantially reduces the number of parameters. The parameter count is proportional to data movement and hence energy consumption. For the liver disease experiment, a single hidden layer digital NN for the original data includes 40000 parameters. In contrast, a ridge classifier with the experimental optical frontend processing results in similar accuracy but uses only 784 parameters. In our framework, the pulse energy is around 50 pJ which is enough for computing by a single pulse with modern sensors[25,26]. Electronics limit our computation speed at input encoding and readout. For proof-of-principle, we used an SLM-based pulse shaper with a 10 Hz refresh rate. However, since our waveguide is on a LN on insulator, the input could be modulated on chip with GHz speed using Mach Zehnder modulators[27,28], already used in a telecommunication network. Furthermore, commercial integrated WDMs could be adapted as a spectral pulse shaper.

The waveguide's spatial domain could provide additional dimensions to encode more input features, which can be used to scale up. For example, evanescently coupled multimode waveguides could be used to increase input size. Therefore, high-speed modulators and multimode waveguides could potentially be used in large data pipelines.

IntSOLO relies on two nonlinear effects: $\chi^{(2)}$ and $\chi^{(3)}$. The optical beams generated by each effect interact with each other during propagation and generate a rich set of optical frequencies. Further analysis and tuning of these two effects may open up unprecedented nonlinearity opportunities. The length of the waveguide is 14 mm, which can be decreased further either by increasing pump power or shortening the pulse duration. The essential point is to obtain enough optical nonlinearity, which is proportional to the product of interaction length and peak power. Kerr frequency combs on LN have already been demonstrated using 300 mW continuous laser[29]. Another exciting direction may be using ring resonators to decrease the required optical power for a compact ultra-low power implementation.

Optical nonlinearity depends on coupled power in the waveguide. We observed instability due to coupled power fluctuation, which is pronounced in our experimental setup because of the long path length of the free space pulse shaper (see Supplementary Discussion 5), sub-micron waveguide cross-section and off-the-shelf components. Pulse shaper could also be implemented using multimode fiber and its intermodal dispersion. Encoding data into different spatial modes could provide temporal optical signals after propagating a few centimeter fibers[30]. In conclusion, this study may provide a path for all-optical information processing for machine learning on a chip.

## Methods

### Data encoding and Feeding into LN waveguide

Data is encoded onto the spectrum of a femtosecond laser using a pulse shaper. We used a spatial light modulator (SLM) based technique for the pulse shaper adapted from ref[31]. It is a quasi 4-f setup in which the pulse is modulated in the frequency domain. The pulse is spectrally dispersed across the horizontal dimension of an SLM using a diffraction grating and focused horizontally with a cylindrical lens. A phase function applied for each frequency is expressed in Equation-1. Where A stands for amplitude and B for phase of a given frequency. A schematic of our optical setup is shown in Supplementary Figure-1. The vertical dimension is used as a programmable amplitude modulator. In the vertical direction, a binary phase grating having a period of 12 pixels ($2z_0$) is used. Amplitude modulation is performed by changing the phase depth of the grating. We use a tunable Chameleon Ultra II Ti: Sapphire laser, which provides 140fs pulses in a wavelength range of 680-1080nm. Our pulse shaper is designed for 950nm. A Holoeye Pluto 2.1 phase only SLM is utilized which has 1920 pixels in horizontal with a pitch of 8 $\mu$m. The unmodulated original pulse's spectral full width at half maximum (FWHM) is roughly 9nm. This corresponds to 500 pixels onto the SLM of the pulse shaper. We measured the two-photon autocorrelation of pulses before and after the pulse shaper to guarantee that the fs pulse is preserved (Supplementary Figure-3). Autocorrelation is performed using a commercial Carpe autocorrelator. We observed 12 fs of broadening, which is expected because of optical components.

$$P(z,\omega) = \left[ rect\left(\frac{z}{2z_0}\right) \cdot e^{i\left(A(\omega) rect\left(\frac{z}{z_0}\right) + B(\omega)\right)} \right] * \sum_{n=-\infty}^{+\infty} \delta(z - 2z_0 n) \quad (1)$$

Datasets are preprocessed before sending to SLM. First, each feature of the dataset is scaled to [0,1]. Secondly, each data sample is arranged to fit a window of SLM and FWHM. If the length of the features is less than 500, features are repeated until the vector length becomes 500. After, the data vector is copied to fill the SLM. Depending on experiments, modulation depth and period can be tuned. For example, if data includes high-frequency features that may cause less average output power due to more diffraction to higher orders. In this case, increasing the modulation period tempers power loss sacrificing dynamic range.

Our waveguides have submicron size. The beam diameter after the pulse shaper is approximately 5 mm. We coupled the pulse shaper output into the LN waveguide with a 60x microscope objective with NA of 0.85. Another microscope objective is used to image the chip from the top which allows to visualize the intensity of the second harmonic generation for fine tuning the coupling. Coupling enough power is crucial for nonlinearity generation. Furthermore, generated second harmonic signal is also exploited to maximize nonlinearity during experiments.

### Digital classification

The final classifier layers are implemented using the scikit-learn python package on a computer. Output spectrums are collected using a spectrometer (Ocean Optic HR 4000). The multimode fiber collecting the light for spectrometer is aligned directly to the waveguide's output as close as possible. For experiments, we first encoded the dataset one by one on pulses, which are sent through the waveguide and then recorded corresponding output spectra. After, we clipped and combined the portions of the spectra around the second harmonic and pump wavelengths. This produces 196 points in our experimental setup. That is, each sample is mapped to a vector of 196 scalar values. This mapped dataset is called a transformed dataset. Then various classification methods are applied to the transformed dataset. The transformed dataset is normalized (zero mean and unit standard deviation) across wavelengths for all classification techniques. 30% of the dataset is used for the test and random train-test split is used 30 times, The average is then reported. The Adam optimizer is used for neural network with a learning rate of $10^{-5}$. Confusion matrices are normalized row-wise. The liver disease dataset is composed of an imbalanced dataset. Therefore, synthetic minority oversampling technique (from scikit-learn) is utilized to balance. For the original dataset (dataset before the optical transform), we up sampled the features so that number of features in the original (input) and optically transformed are the same. This is important because the number of features defines the parameter counts in the model and the parameter count is proportional to model capacity. In this work, experimentally 196 wavelengths are used at the output. Therefore, the original features are up sampled to 196.

### LN Chip

The $LiNbO_3$ waveguide is fabricated in a 400 nm thick thin-film $LiNbO_3$ on 2 μm $SiO_2$ on 500 μm Si wafer with an x-cut crystal orientation as previously described in ref[18]. The fabrication uses standard electron-beam lithography with HSQ resist and an optimized ICP/RIE dry-etching with Ar ions. The remaining mask and dry-etching residuals are removed by a subsequent wet-etching step in a buffered oxide etch (BOE) and potassium hydroxide (KOH). The waveguide has a length of 14 mm and a top width of 750 nm, ridge-height of 250 nm, and remaining thin-film thickness of 150 nm, and a sidewall angle of 60°. After dicing the input facets of the waveguides are mechanically lapped and polished to enhance light coupling efficiency.

### Acknowledgment

We acknowledge support for nanofabrication from the Scientific Center of Optical and Electron Microscopy (ScopeM) and from the cleanroom facilities BRNC and FIRST of ETH Zurich. This work was supported by the Swiss National Science Foundation, Grant No. 179099, and by the European Union's Horizon 2020 Research and Innovation Program from the European Research Council, under Grant Agreement No. 714837 (Chi2-nano-oxides).

## Data availability

The data and findings of this study are available from the corresponding author upon reasonable request.

## Author Contributions

M.Y., D.P. and C.M. conceived the project. M.Y. and I.O. performed simulations and experiments. F.K. and M.R.E fabricated the chip. R.G., D.P. and C.M. supervised the project.